\documentclass[preprint2,amsmath,amssymb]{aastex}

\usepackage{graphicx}
\usepackage{latexsym}

\begin{document}


\title{Gravitational cooling of self-gravitating Bose-Condensates}
\author{F. Siddhartha Guzm\'an}
\affil{Instituto de F\'{\i}sica y Matem\'{a}ticas, 
Universidad Michoacana de San Nicol\'as de Hidalgo. Edificio C-3, Cd. 
Universitaria. C. P. 58040 Morelia, Michoac\'{a}n, M\'{e}xico.}

\and
\author{L. Arturo Ure\~{n}a-L\'{o}pez}
\affil{Instituto de F\'isica de la Universidad de Guanajuato, 
A.P. 150, C.P. 37150, Le\'on, Guanajuato, M\'exico.}


\begin{abstract}
Equilibrium configurations for a self-gravitating scalar field with 
self-interaction are constructed. The corresponding
Schr\"odinger-Poisson (SP) system is solved using finite differences
assuming spherical symmetry. It is shown that equilibrium
configurations of the SP system are late-time attractor solutions for
initially quite arbitrary density profiles, which relax and virialize
through the emission of scalar field bursts; a process dubbed
gravitational cooling. Among other potential applications, these
results indicate that scalar field dark matter models (in its
different flavors)  tolerate the introduction of a self-interaction
term in the SP equations. This study can be useful in exploring models
in which dark matter in galaxies is not point-like.
\end{abstract}



\section{Introduction}
\label{sec:intro}

One of the most fundamental problems in modern Cosmology is the nature
of dark matter. This problem itself has provoked an intense
theoretical search for realistic candidates able to drive the
formation of the cosmological structure we observe today \cite{sahni2004}. 
Among others, there are some candidates that have for long been considered
as exotic; a paradigmatic example is the case of scalar fields, which
are the simplest particles we can think of \cite{matoslurena2004, 
sahni2000}. However, no fundamental scalar field has ever been detected. 

Nevertheless, theoretical research on scalar candidates is a very
interesting field in Cosmology nowadays, and many ideas have been
proposed. One principal idea is that scalar fields can form Bose
condensates, and then are able to form
structure \cite{hu, arbeyetal, matoslurena2001}. The quest for the kind
of objects that are formed by scalar fields has been investigated in
detail using numerical tools. The relativistic regime has been studied
in \cite{seidel94, balakrishna98, alcubierre2002, alcubierre2003}, for
both complex and real scalar fields.

Scalar fields are an example in which dark matter is not point-like,
hence no conclusions can be extracted from the widely studied N-body
simulations for dark matter particles. Thus, we need to understand
their properties in order to decide whether such properties could be
useful to explain what we see in galaxies. On of the most attractive 
features of the scalar field model is that the construction of equilibrium 
configurations demands a smooth density profile at the origin, which is an 
advantage over the point-like dark matter particle models which have shown 
to be cuspy.

In general, complex scalar fields can form stable equilibrium configurations
called \textit{boson stars}, that are globally regular and whose
energy density is time-independent, for a recent review
see \cite{schunck}. Real scalar fields also have equilibrium
configurations, that were discovered in \cite{seidel91}, and are called
\textit{oscillatons}. The latter are also globally regular, but are
fully time-dependent. As for their stability, it seems to be 
quite robust as far as numerical evolutions is
concerned \cite{seidel91,alcubierre2002,alcubierre2003}, but it may be
that they are only long-lived \cite{d-page}. 

An important part of any scalar field model of dark matter is
  the \textit{scalar potential} $V(\phi)$, which encodes in itself the
  self-interactions of the scalar field other than gravitational. The
  most popular scalar potential, and also the most studied in
  Cosmology, is the quadratic one, $V(\phi)=(m^2/2) \phi^2$ (or
  $V(|\phi|)=m^2 |\phi|^2$ if the scalar field is complex), where the
  parameter $m$ is identified as the mass of the boson particle.

On the other hand, the inverse of the mass, i.e. the
  Compton length of the boson $\lambda_C = m^{-1}$ (in units in which
  $\hbar =c=1$), becomes the natural unit of distance of scalar
  configurations. Actually, the most compact scalar objects have a
  radius of a few Compton lengths. In addition, the boson mass also
  establishes the mass scale of the compact object, which is of the
  order $\sim m^2_{Pl}/m$, where $m_{Pl}$ is the Planck
  mass. Thus, a light scalar mass may provide of very massive
  objects\cite{seidel91,harri1,alcubierre2003,schunck,fsglau2004}.
  However, scalar potentials more complicated than the quadratic one
  can provide of extra parameters which can add new features to
  oscillatons and boson stars.

One simple instance of such more general scalar potential is
  that containing an extra quartic coupling in the form $V(\phi)=(m^2/2)
  \phi^2+\lambda_{int} \phi^4/4$, see
  \cite{wasserman,balakrishna98,arbeyetal} and references therein. In
  this case we speak of a self-interacting scalar field\footnote{The
  quadratic potential is also called \emph{free} potential, as the
  scalar field represents in this case a free relativistic particle of
  mass $m$ with total energy $E^2 = p^2 + m^2$.}.

The influence a quartic self-interaction can have in the
  dynamics of a cosmological \textit{complex} scalar field has been
  widely studied in the literature, for recent developments
  see\cite{schunck,arbeyetal}. However, there are also some changes
  with respect to the free case. The self-interaction term can become
  important in situations in which the scalar field has large amplitude, 
which
  may happen typically in the early universe, and in the interior of
  compact scalar objects. For the purposes of this paper, it should be
  mentioned that a quartic term makes, for a given scalar field
  strength, a boson star larger and more massive than in the free
  case\cite{wasserman,balakrishna98,guzman2006}.

On the other hand \textit{oscillatons} with self-interacting terms in
  the scalar potential have not been studied yet. The reason for this
  may be an intrinsic difficulty in dealing with time-dependent
  relativistic configurations.

Nevertheless, a curious point is that the Newtonian regime is the same
for boson stars and oscillatons. The relativistic EKG equations in
this regime are equivalent to the so-called Schr\"odinger-Poisson (SP)
system, see \cite{seidel94,harri1,fsglau2003,fsglau2004} and
references therein. 

For a free scalar potential, the Newtonian equilibrium configurations can
be as large as a galactic halo if the mass of the boson is very light
\cite{arbeyetal,fsglau2003}. That the Newtonian regime is the
  astrophysically interesting one can be seen if one calculates the
  Compton length associated to a (very light) boson mass of order
  $10^{-23}$ eV; this is $\lambda_C \sim 1$
  pc\cite{arbeyetal,sahni2000,matoslurena2001,hu}. From the point
  of view of particle physics, this boson mass is extremely
  small. Therefore, such a light boson will form \emph{relativistic}
  objects which are small compared to typical galactic
  scales. However, in the \emph{Newtonian} regime, gravity is weaker
  and allows the formation of much larger scalar objects.

Within the Newtonian regime, there are also equilibrium solutions for 
different number of nodes of the wave function; solutions with nodes are 
called excited Newtonian boson stars and are classified according to the 
number of nodes the wave function has. Zero-node states are called 
\emph{ground} states, and configurations with nodes are generically
called \emph{excited states}. Such excited states were also considered in 
the past as candidates for dark matter halos in galaxies \cite{koreanos}. 
They have the nice property of providing nearly flat rotation curves if 
the number of nodes is sufficiently large, see the example shown 
in \cite{arbeyetal}. Unfortunately, those excited boson stars are not 
stable, and they decay into a ground state in a time much smaller than the 
actual age of the universe \cite{fsglau2003,fsglau2004}; the attractor 
behavior of the ground state configurations is shown below in this paper 
for initially excited state equilibrium configurations. This fact 
rules out the possibility of excited boson stars as dark matter halos; 
nevertheless, they could still play the role of transient states in the 
evolution of scalar field configurations. 

It is interesting to note that the SP system has been widely studied
in other fields out of cosmology, see the interesting works
in \cite{dale-phd, dale-choi, giovan2001}. Also, they have been proposed to
ameliorate the quasar's redshift problem in \cite{svid2004}.

In this paper we want to study the case of scalar configurations with
a self-interaction term in the Newtonian limit, that is, we shall
study the SP system including a quartic self-interaction term. In
doing this, we will learn about the properties that both a boson star
and an oscillaton have in the weak field limit.

Scalar field configurations with a quartic
  self-interaction have been proposed before to be the dark
  matter in galaxies, see \cite{arbeyetal}. But, it is
  necessary to understand the dynamical properties of such
  configurations. In this sense, this paper extends the research
  already done in\cite{fsglau2004} for the (free field) SP system.

Assuming spherical symmetry, the Schr\"odinger-Poisson (SP) 
system of equations to be solved reads
\label{sp}
\begin{eqnarray}
i\partial_\tau \psi &=& -\frac{1}{2x} \partial^2_x (x\psi) + U \psi    
+ \Lambda |\psi|^2\psi
\label{schroedinger}\\
\partial^2_x (xU) &=& x \psi \psi^\ast . \label{poisson}
\end{eqnarray}
We have used the dimensionless variables $\tau = mt$, 
$x=mr$, where $t$ and $r$ are the physical time and radial coordinates, 
respectively, and $m$ is the mass of the boson.

The wave function $\psi$ is coupled to its own gravitational potential 
$U$, and the constant $\Lambda$ represents the $s$-wave scattering
length in the Gross-Pitaevskii approximation for Bose
condensates \cite{gross,pitaevskii}. 

It is known that the SP set of equations without self-interaction has
a fancy scaling property \citep{fsglau2004} that simplifies the
numerical research. It is possible to establish a similar scaling
property when a self-interaction term is included. That is,
Eqs.~(\ref{sp}) are invariant under the transformation
\begin{equation}
\{\tau,x,U,\psi,\Lambda \}
 \rightarrow
\{\lambda^{-2}\hat{\tau},
\lambda^{-1}\hat{x},
\lambda^{2}\hat{U},
\lambda^{2}\hat{\psi},
\lambda^{-2}\hat{\Lambda} \} \label{eq:scaling}
\end{equation}

\noindent with $\lambda$ an arbitrary parameter. Unfortunately, there
is a main difference with respect to the non-interacting case. In the
latter, the solution of the SP assuming $|\psi(0)|=1$ generates the
whole and unique branch of possible equilibrium solutions.

For the new SP equations~(\ref{sp}), it is necessary to find first all
possible solutions that belong to a \emph{single} value of the self-interaction
coefficient $\Lambda$. If we now scale the latter using a certain
$\lambda$, we can generate a full \emph{new} branch of solutions for
$\hat{\Lambda}$ out of the one belonging to the old $\Lambda$. Though
not as easy as in the free case, the scaling
transformation~(\ref{eq:scaling}) will still help us to understand the
evolution of a scalar field configuration under an arbitrary value of
$\Lambda$.

There is also a complete set of physical quantities that are well
defined in the Newtonian limit. They obey  the following scaling 
transformation

\begin{equation}
\{\rho,N,K,W,I \} \rightarrow
\{\lambda^{4}\hat{\rho},
\lambda \hat{N},
\lambda^{3} \hat{K},
\lambda^{3} \hat{W},
\lambda^{3} \hat{I} \} \, , \label{props}
\end{equation}

\noindent where $\rho$ is the density of probability, $N$ is the integral
of $\rho$ over the whole space, $K$ and $W$ are the expectation values
of the kinetic and gravitational energies, and $I$ is the expectation
value of the self-interaction energy. The above quantities will help
us to follow the evolution of an arbitrary scalar field configuration and 
establish criteria about virialization and equilibrium,

This paper is organized as follows. In Sec.~\ref{sec:ivp}, we present
the solution to the initial value problem that will be the equilibrium
configurations of the SP system. In Sec.~\ref{sec:numerics}, we
describe the numerical methods and the boundary conditions used for
the time evolution of the SP system. After that, in 
Sec.~\ref{sec:cooling}, we describe what we mean by virialization of 
non-equilibrium configurations and the late-time attractor behavior of the 
sequence of equilibrium configurations. Finally, in 
Sec.~\ref{sec:conclusions}, we draw some comments and conclusions.

\section{Equilibrium configurations}
\label{sec:ivp}

For equilibrium configurations, we assume that the wave function is of the
form $\psi(\tau,x) = e^{i\omega \tau} \phi(x)$, where $\omega$ is a
free parameter. This assumption implies that the density of
probability $\rho$ and the gravitational potential $U$ are
time-independent, whereas the wave function evolves harmonically in
time. Eqs.~(\ref{schroedinger}) and~(\ref{poisson}) become
\begin{eqnarray}
\partial^{2}_{x}(x\phi)&=& 2 x (U-\omega) + 2\Lambda
|\phi|^2\phi \, , \label{sch-spherical} \\
\partial^{2}_{x}(xU)&=& x\phi^2 \, . \label{poi-spherical}
\end{eqnarray}

\noindent The above system has to be solved under the condition of
regularity at the origin $\phi(0)=\partial_x \phi(0)=0$, and isolation
$\phi(x\rightarrow \infty) = 0$; we will also demand that
$U(\infty)=0$. These boundary conditions determine in a unique manner
the values of $U(0)$ and $\omega$, which are the only free parameters
of the solution.

\begin{figure*}
\includegraphics[width=8cm]{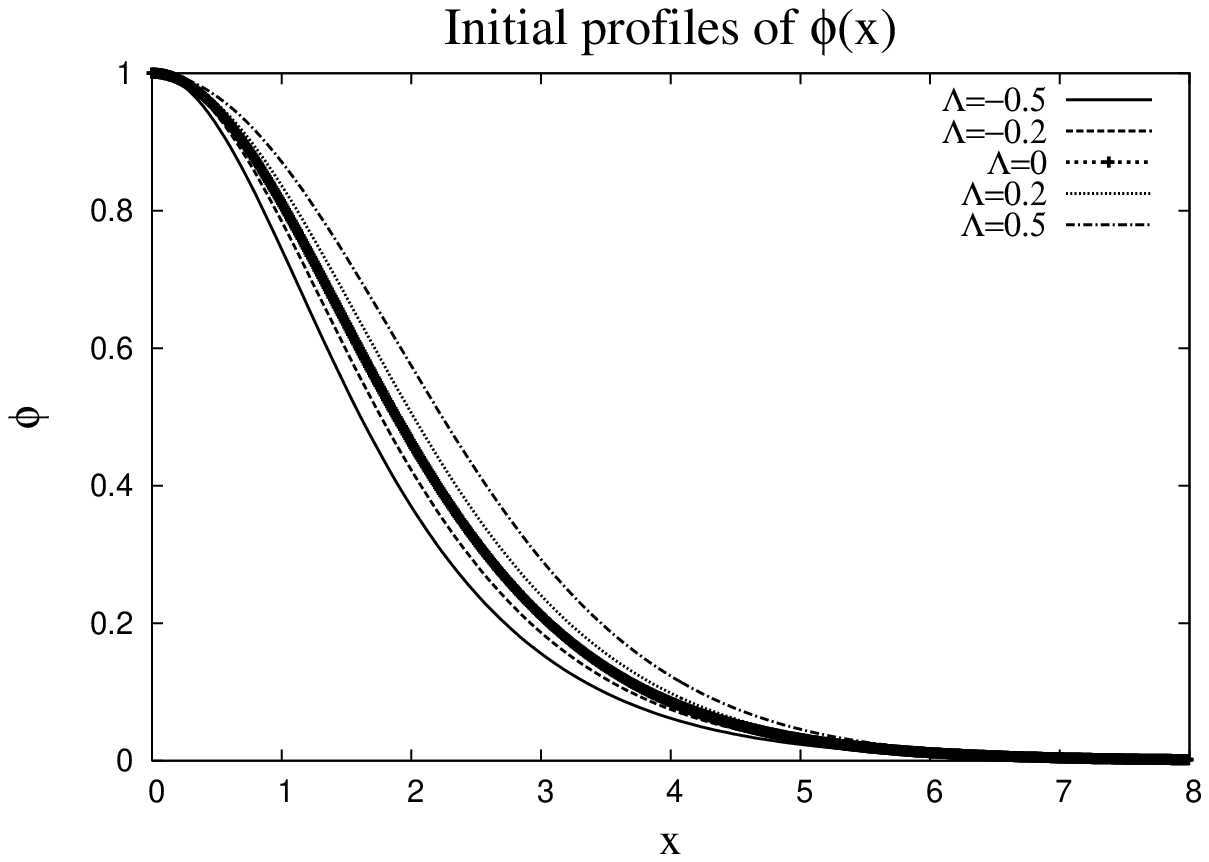}
\includegraphics[width=8cm]{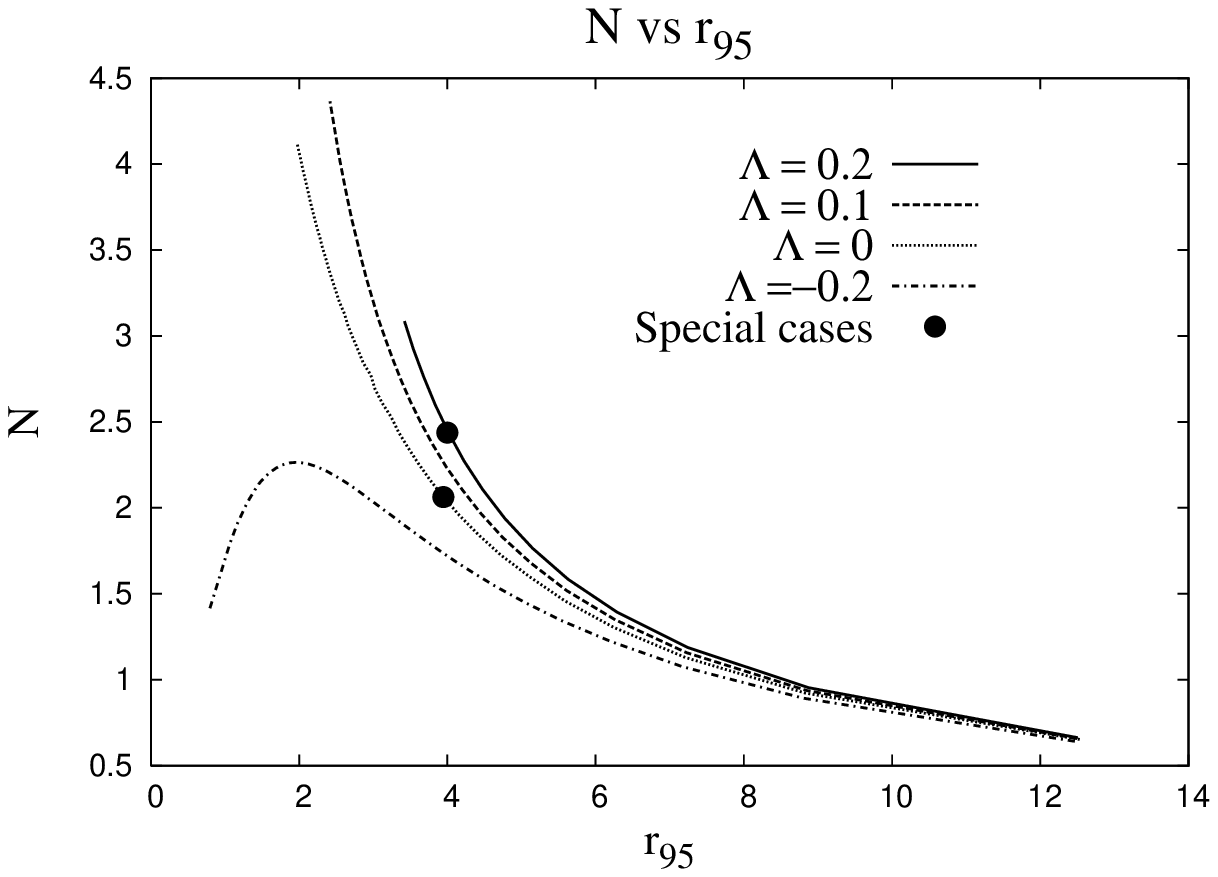}
\caption{\label{fig:plotone} (Left) Profiles $\phi(x)$ of equilibrium 
solutions of the SP for different values of  the self-interaction 
coefficient $\Lambda$. As expected, the larger the coefficient $\Lambda$ 
the more massive the equilibrium configuration is. (Right) 
Sequences of equilibrium configurations for different values 
of $\Lambda$. Each point in the curves represents a solution of the 
initial value problem of a total number of particles $N$ and $95$\%
radius $r_{95}$ (the radius inside which $95$\% of the total mass is
contained in). In this plot it is manifest that the bigger the $\Lambda$ 
the less compact a configuration is. The filled circles indicate two 
configurations we use as tests for our numerical implementation.} 
\end{figure*}

We made a numerical routine that solves Eqs.~(\ref{schroedinger})
and~(\ref{poisson}) using a shooting procedure that tunes the values
of $\omega$ and $U(0)$ for a given central value $\phi(0)$. In
Fig.~\ref{fig:plotone}, we show the profiles of the initial wave function
for some positive and negative values of the self-interaction
coefficient $\Lambda$. Also shown in there are the branches of
equilibrium configurations for four different values of 
$\Lambda$. As expected, for $\Lambda \ge 0$ there is nothing like a 
maximum indicating an unstable branch, which is always present (and 
typical) in relativistic boson configurations
\cite{wasserman,balakrishna98,guzman2004}. 

\section{Numerical evolution}
\label{sec:numerics}

In order to numerically evolve the SP system, we make a discretization
of space and time, and approximate all derivatives using second
order accurate finite differencing. The SP system is evolved one time
step $\Delta \tau$ using Eq.~(\ref{schroedinger}) to obtain a new wave
function, and then we solve Eq.~(\ref{poisson}) to find the
corresponding (new) gravitational potential. 

We use an explicit time integrator to solve Eq.~(\ref{schroedinger}), as
opposed to the fully implicit method used and described
in\cite{fsglau2004}, where the problem of evolution was reduced to a
linear system of equations. In the present case, due to the non-linear
term in the Hamiltonian of Eq.~(\ref{schroedinger}), the reduction to
a linear system of equations is not that simple. 

In any case, independently of the numerical method used to integrate
in time, we demand the evolution operator to preserve the number of
particles for an \textit{equilibrium configuration} $N=\int |\psi| x^2
dx$\footnote{In full units, the total mass is given by
  $M=(m^2_\textrm{Pl}/m) N$, where $m_\textrm{Pl}$ is the Planck
  mass.}. With a modified version of the iterative Crank-Nicholson
evolution method \cite{teukolsky00a}, we were able to reproduce the
results found with the implicit method for the linear case, and
confirm that the evolution was mass-preserving for a wide range of
values of $\Lambda$.

\begin{figure*}
\includegraphics[width=8cm]{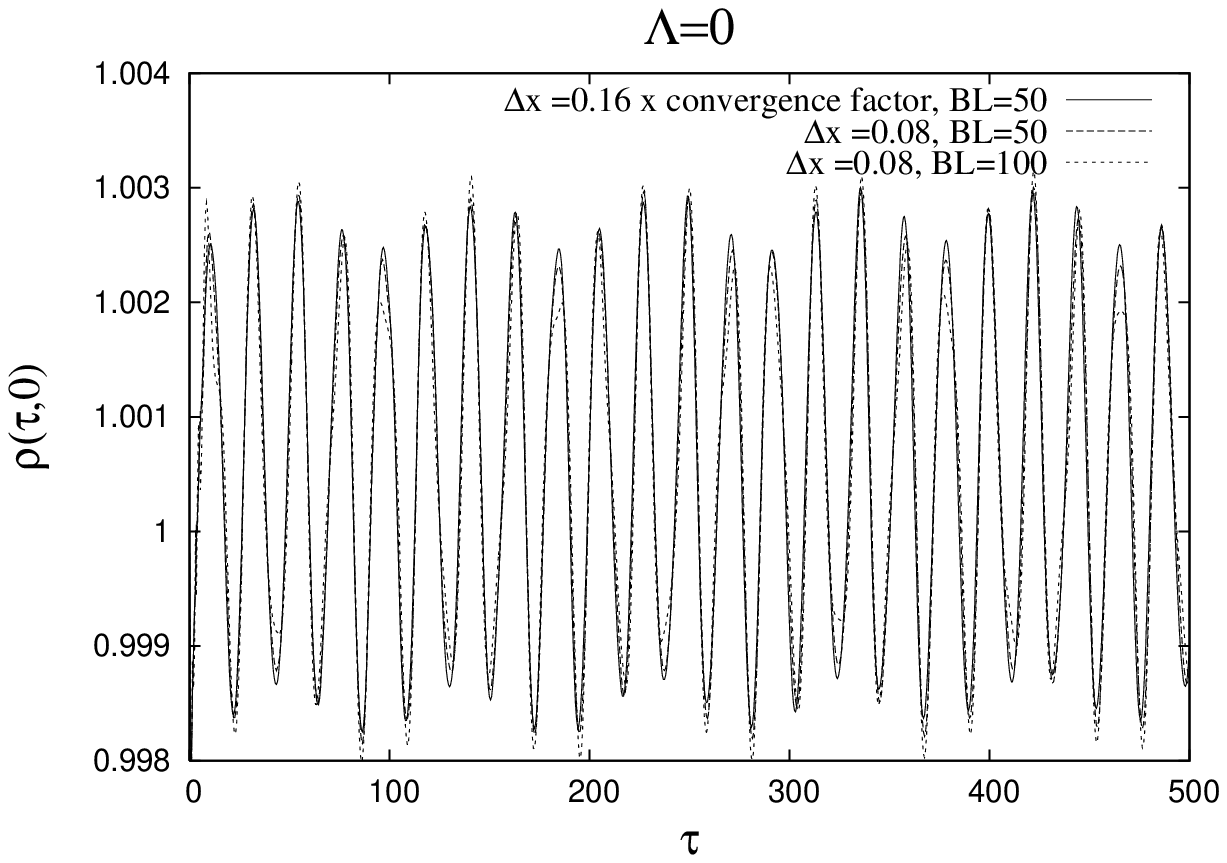}
\includegraphics[width=8.5cm]{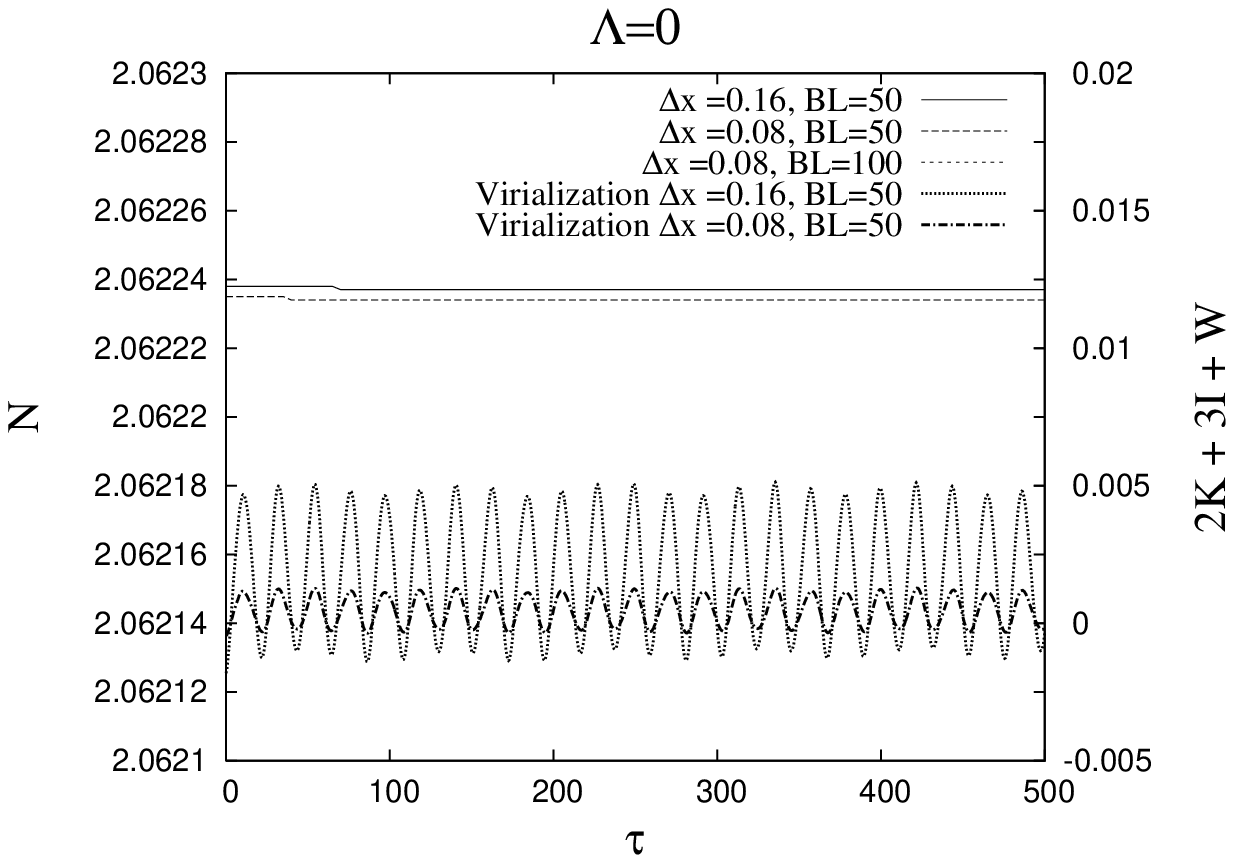}
\includegraphics[width=8cm]{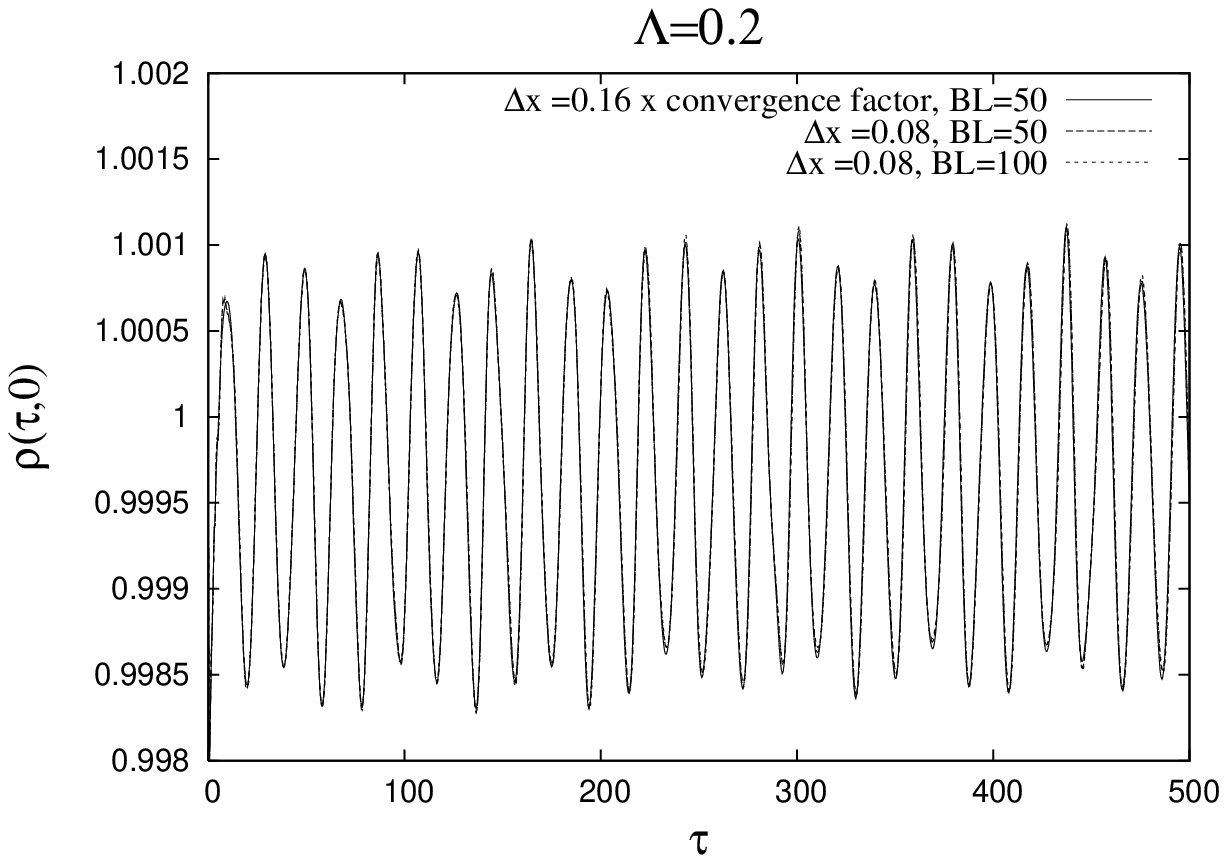}
\includegraphics[width=8.5cm]{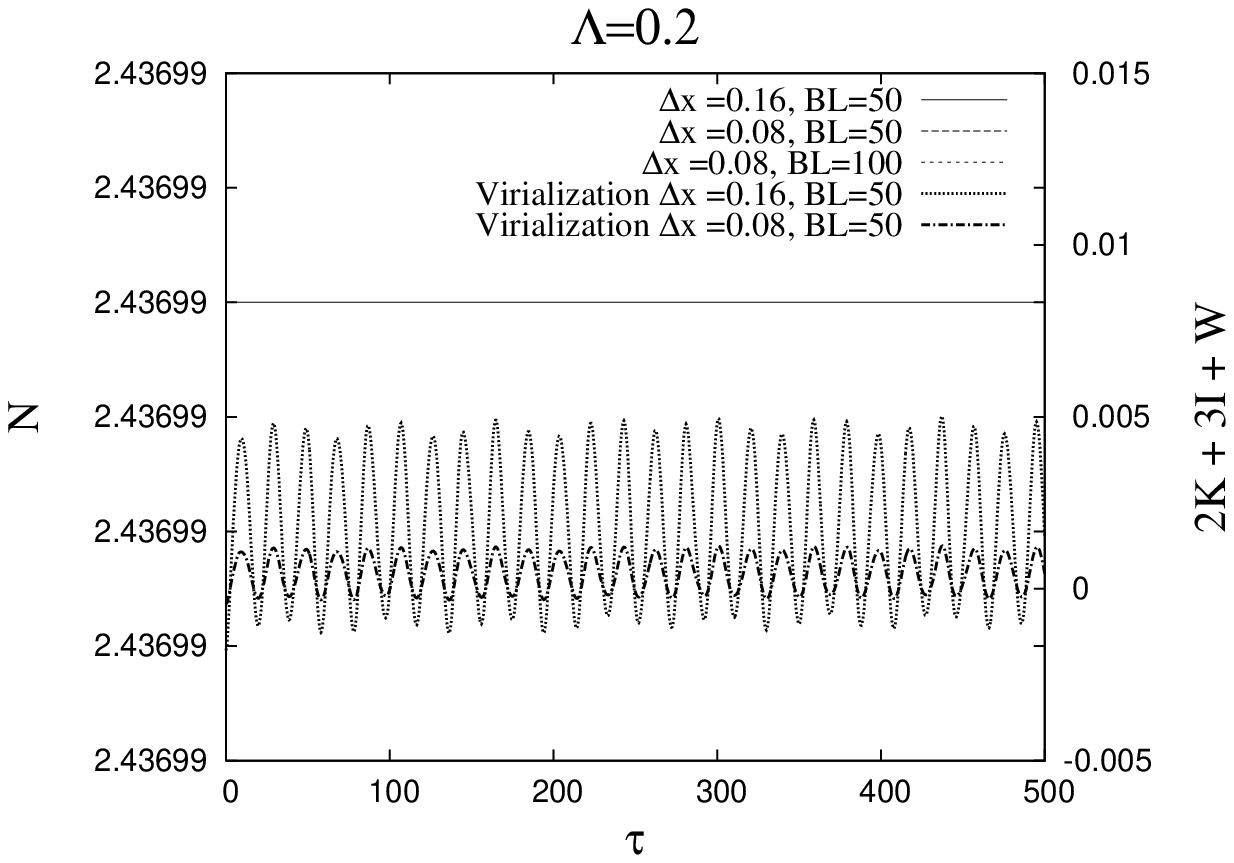}
\caption{\label{fig:plottwo} Evolution of equilibrium configurations 
with different values of $\Lambda$. \textit{Figures on the left}. For
each case, the central value of $\rho(\tau,0)$ is shown for two
resolutions: $\Delta x = 0.08, 0.16$ with the boundary located at
$x_B=50$, and a third case with the boundaries located at $x_b=100$
and resolution $\Delta =0.08$. The curve corresponding to the coarse
resolution has been scaled as $(\rho_{0.16}(0) - 1)/4 + 1$, and the
fact that it lies upon the value found for $\rho_{0.08}$ indicates
second order convergence of the solutions. Hence, we are running the
simulations in the convergence regime of our discretization. The fact
that the density found for $\Delta x =0.08$ with the boundaries at two
different locations lies upon each other, indicates that the results
are independent of the location of the boundary. \textit{Figures on
  the right} We show the evolution of the number of particles, and the
value of the virialization quantity $2K+3I+W$, which in the continuum
case should be zero. We see that its values are four times smaller if
the spatial resolution is doubled. This clearly indicates second order
convergence of this quantity to zero, which in turn means that we can
recover what is expected in the continuum limit.} 
\end{figure*}

{\it Boundary conditions}. At every time step, Eq.~(\ref{poisson}) is
integrated inwards, and thus we applied the following boundary
condition at the two outermost points of the numerical grid
\begin{eqnarray}
U(x_{n-1}) &=& -N(x_{n-1})/x_{n-1}  \, , \label{bcU1}\\
U(x_n) &=& -N(x_n)/x_n  \, .\label{bcU}
\end{eqnarray}

\noindent where $n$ labels the outermost point of the numerical
domain. The whole profile of $U(x)$ is then found using a sixth order
accurate Numerov algorithm \cite{cphysics}.

As it is discussed in \cite{fsglau2004}, the boundary
conditions (\ref{bcU1}) and (\ref{bcU}) are equivalent to impose 
the condition $|\psi(\tau, x_n)| \rightarrow 0$ on the wave
function; hence, we are forcing the system to remain in the
computational domain. Because we want to evolve systems out of
equilibrium and allow the flow of particles out of the numerical
domain, we implemented a \textit{sponge} over the outermost points of
the grid, which consists in adding an imaginary potential $V_j(x)$ to
the Schr\"odinger equation (see \cite{fsglau2004,israeli}). The
expression we use for the sponge profile is
\begin{equation}
V_j = -\frac{i}{2} V_0 \left\{ 2 + \tanh \left[(x_j-x_c)/\delta 
\right] - \tanh \left( x_c/\delta \right) \right\} \, , \label{imagpot}
\end{equation}

\noindent which is a smooth version of a step function with amplitude 
$V_0$, centered in $x_c$ and width $\delta$. the minus sign warranties
the decay of the number of particles at the outer parts of our
integration domain, that is, the imaginary potential behaves as a sink
of particles. It is also worth noticing that no term related to the
self--interaction appears in this conservation equation since we are
assuming $\Lambda$ is real.

{\it Tests.} The obstacles our code must sort out are: i) the
evolution of equilibrium configurations, in which the wave function
oscillates with the definite frequency $\omega$ found in
Sec.~\ref{sec:ivp}, whereas the density of probability and the
gravitational potential should remain time-independent; ii) the
convergence of physical properties of the system~(\ref{props}).

These two tests are shown at once in Fig.~\ref{fig:plottwo}. On the
left hand side, we show the convergence of the central density and the
independence of the evolution on the location of the numerical
boundary. We know that in the continuum limit, these configurations
should evolve keeping $\rho(\tau,0)=1$ for all times. It is shown that
our approximation through finite differences converges to that value
in the continuum limit. 

The central density shows oscillations whose amplitude converge to zero as the
numerical grid is refined. We relate these oscillations to the
response of the system to the intrinsic perturbation induced by the
truncation error of the finite differencing approximation\footnote{In
  \cite{fsglau2004}, we compared this type of oscillations for $\Lambda
  =0$ with a linear perturbation analysis and found excellent
  coincidence with the Fourier analysis of the numerical evolutions. In the
  present case, we deal with a non-linear Schr\"odinger equation, and
  such analysis would say little about the cases with $\Lambda \ne 0$.
  Nevertheless, we already have evidence that the truncation error could be
  responsible for this very regular oscillations.}.
On the right hand side, we show how the evolution operator preserves
the number of particles all along the simulation. 
In a few words,
what we show in Figure \ref{fig:plottwo} is how the evolution
method, the finite differencing and the boundary conditions work
properly all together.

Besides the expected results found for the cases marked with filled 
circles in Fig.~\ref{fig:plotone}, we also tried to evolve configurations 
located to the left of the maximum for $\Lambda=-0.2$. Such configurations 
are very compact, and their evolution resulted in a quick collapse
and in the divergence of the central density $\rho(\tau,0)$. In
principle, this would be the expected behavior for an unstable
equilibrium configuration in the \textit{relativistic} regime. However, we do not account with a clear definition of a collapsed 
object in terms of trapped surfaces of horizon formation during the 
evolution, so that it is not possible to draw accurate statements
within the non-relativistic SP formalism for $\Lambda < 0$. One 
possibility would be to continue the evolution of the system with a 
relativistic code after certain compactness is achieved, however such 
analysis is beyond the scope of this paper. 

\section{Gravitational cooling and virialization}
\label{sec:cooling}

The virialization and stabilization of physical configurations 
evolving according to Eqs.~(\ref{schroedinger}) and~(\ref{poisson})
with $\Lambda=0$ was first proposed in\cite{seidel94}, and revised in
detail in \cite{fsglau2004}. The mechanism for relaxation consists in
the ejection of scalar field bursts, and the tendency of the system to
settle down onto equilibrium configurations. This process was dubbed
\textit{gravitational cooling}. 

The criterion to determine whether or not a system was near an
equilibrium state involves monitoring the virial theorem relation
\begin{equation}
2K + 3I + W = 0 \, , \label{eq:virial}
\end{equation}

\noindent where $K,I,W$ are the expectation values of the kinetic,
self-interaction, and gravitational potential energies, respectively,
for the SP system (see for instance \cite{wang2001}). In
Fig.~\ref{fig:plottwo}, we also show the oscillatory behavior of
the quantity $2K + 3I + W$, whose amplitude converges to zero in the
continuum limit.

\begin{figure*}
\includegraphics[width=8cm]{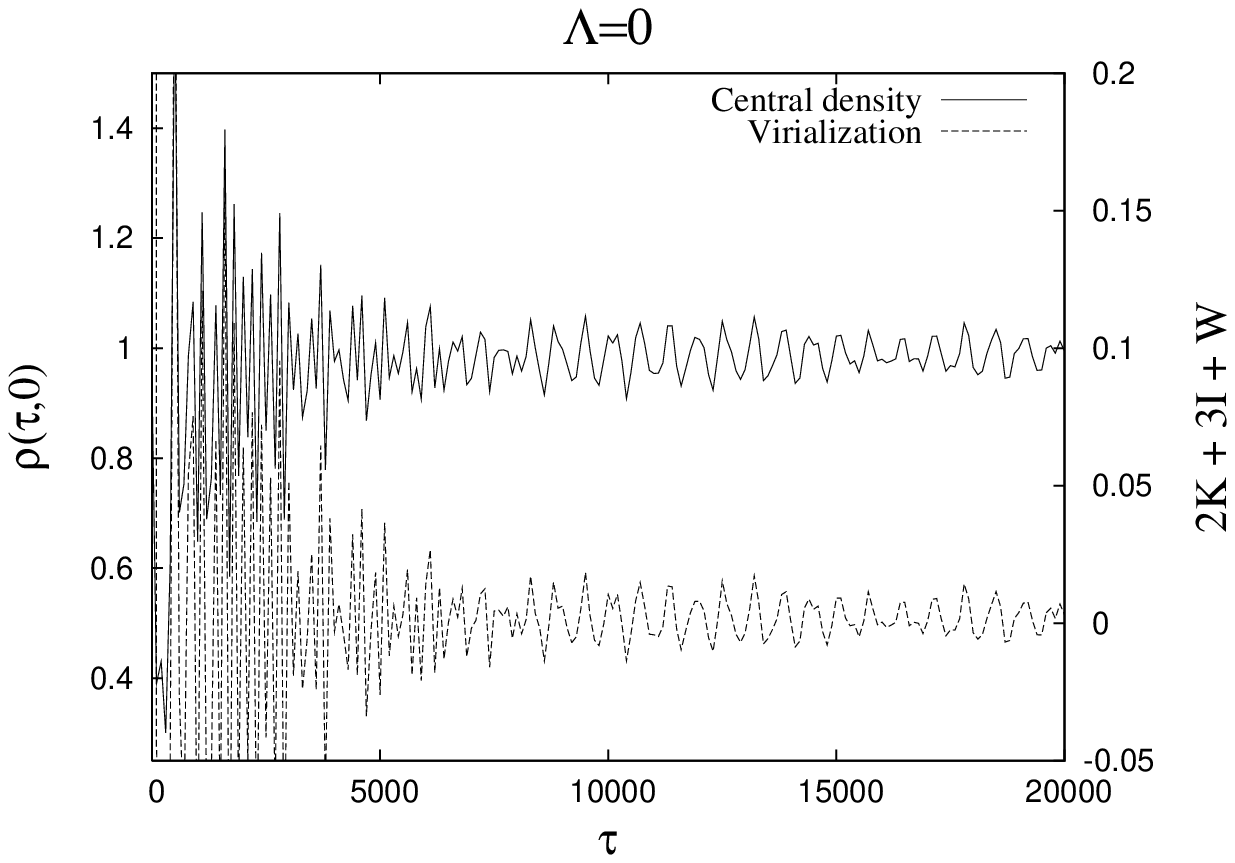}
\includegraphics[width=8cm]{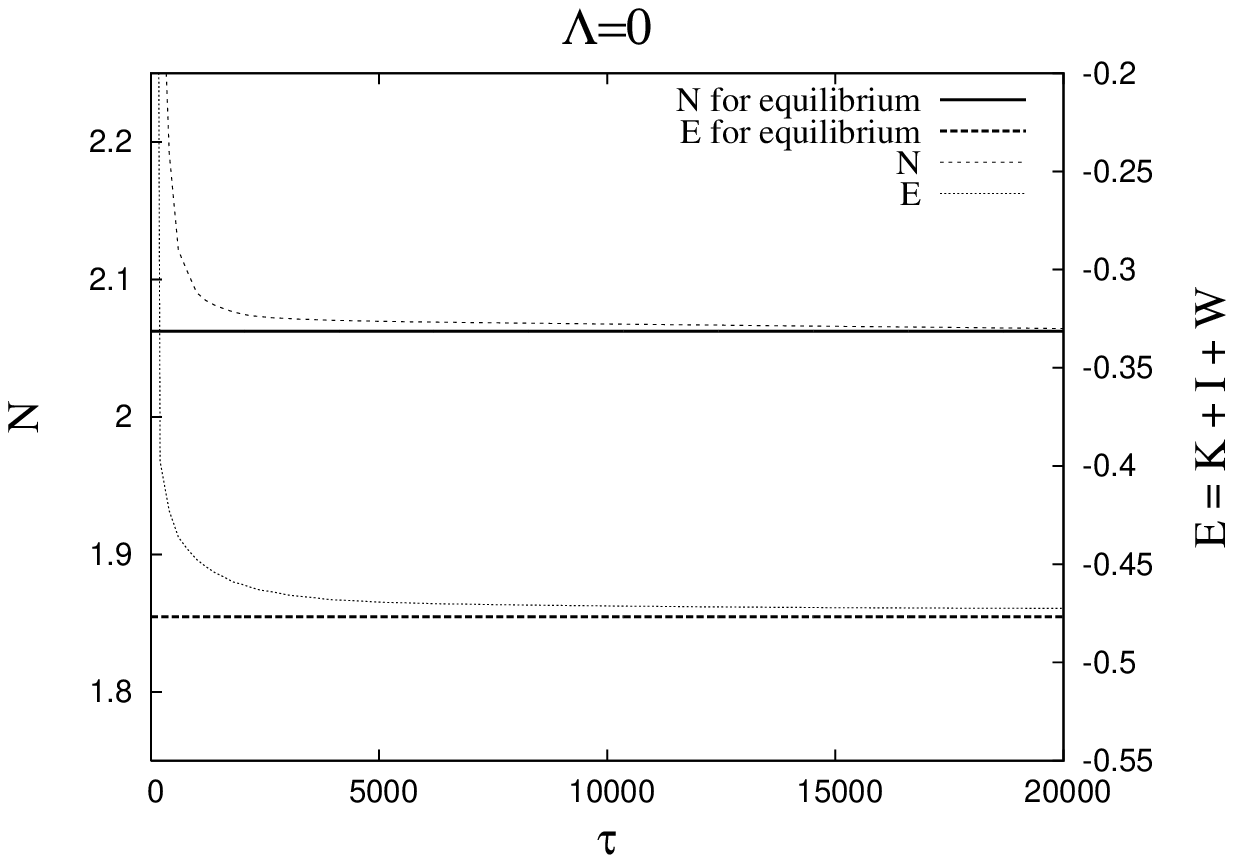}
\includegraphics[width=8cm]{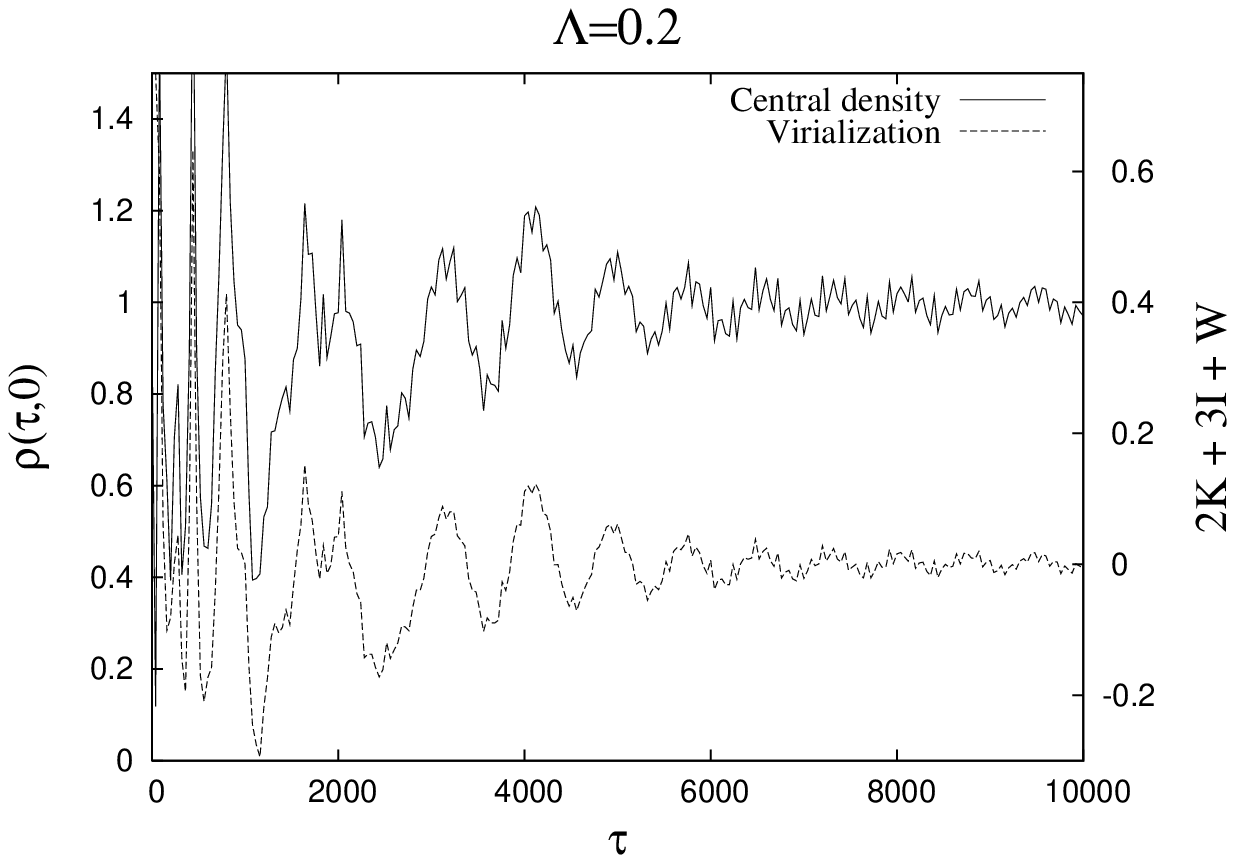}
\includegraphics[width=8cm]{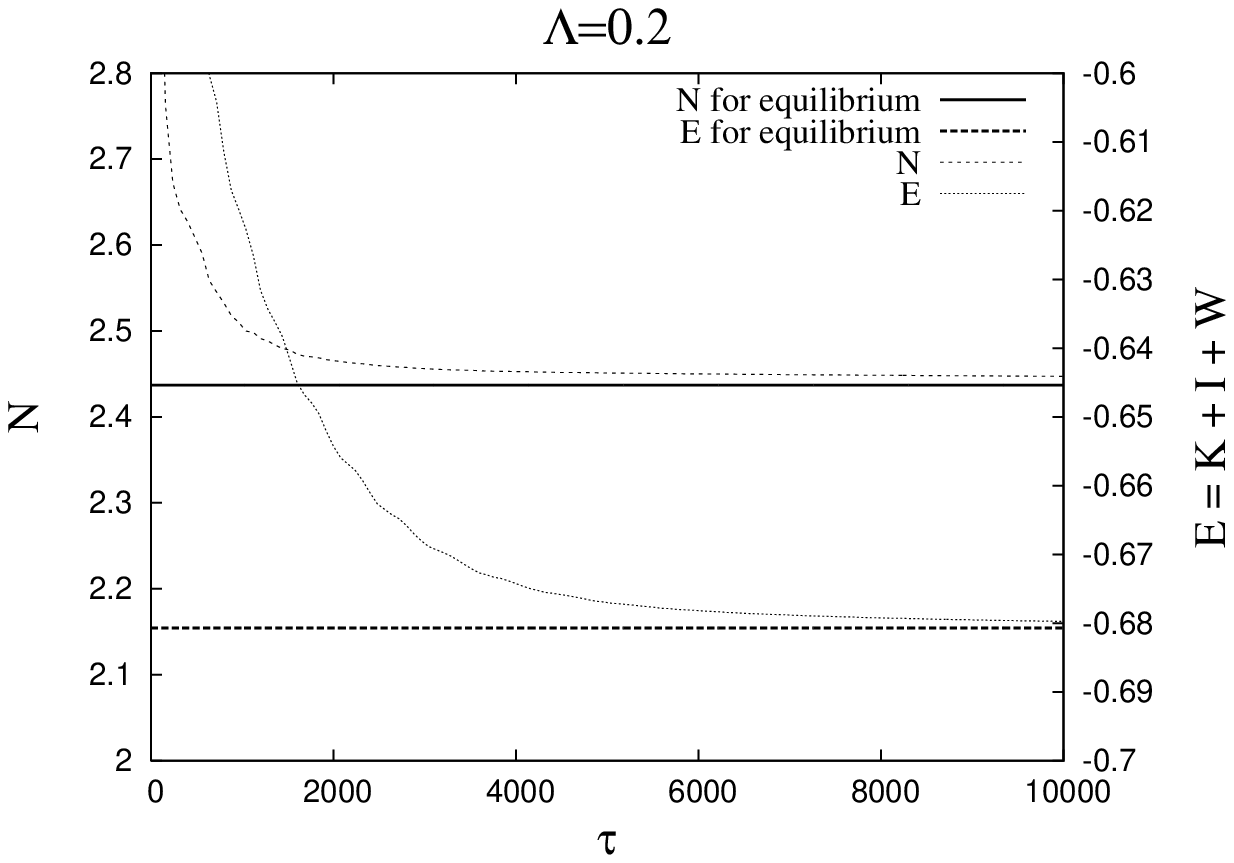}
\caption{\label{fig:plotthree} \textit{Top panel} We show the evolution of 
the 
initial profile $\psi(0,x)=\cos^2(x) e^{-x^2/(3.959)^2}$ with
$\Lambda=0$. The system starts with a high kinetic energy, and after a
relaxation period the central density $\rho(\tau,0)$ stabilizes around
one, whereas the virial relation $2K + 3I + W$ oscillates around
zero. The assymptotic behaviors of the the number of particles $N$ and
total energy are shown on the right plot. The initial mass 
is $M_i=3.617337 \, (m^2_{Pl}/m)$, and during the process around
$42\%$ of it was ejected very quickly. \textit{Bottom panel} This is
the case for $\Lambda=0.2$, with an initial wave function profile
$\psi(0,x) = \cos^2(x) e^{-x^2/(4.085)^2}$. As before, the system
also relaxes and stabilizes around an equilibrium configuration. This
time, the initial mass is $M_i=3.98 \, (m^2_{Pl}/m)$, and the ejected
mass is $\sim$39\%. In both cases the final configurations are nearly 
those shown in Figure~\ref{fig:plottwo}.} 
\end{figure*}

In Fig.~\ref{fig:plotthree}, we show the virialization process for two
different values of $\Lambda$. The initial profile is of the form
$\psi(0,x) = \cos^2(x) e^{-x^2/\sigma^2}$, with $\sigma$ a free parameter. The
reason why we choose such initial profile is that it provides a high
initial kinetic energy, and the process of virialization (if
any) will be evident while calculating Eq.~(\ref{eq:virial}); other 
initial profiles of garden variety have been evolved, however the 
overwarmed initial configurations shown here present in a better way the 
relaxation process. The case $\Lambda=0$ is as simple as those found in 
the past \cite{seidel94,fsglau2003,fsglau2004}, because it is possible to
choose any profile with the confidence that the configuration will
approach a rescaled equilibrium configuration (as shown in
\cite{fsglau2004}), but we want to be sure that the same happens for
$\Lambda \neq 0$.

Our expectations were fulfilled because of the manifest results in
Fig.~\ref{fig:plotthree}. All of the studied cases show a relaxation
process towards one of the equilibrium configurations shown in
Fig.~\ref{fig:plotone}. With these
results, we confirm that the equilibrium configurations in
Fig.~\ref{fig:plotone}, which are solutions of Eqs.~(\ref{sch-spherical})
and~(\ref{poi-spherical}), are indeed late-time attractor solutions.

\begin{figure*}
\includegraphics[width=8cm]{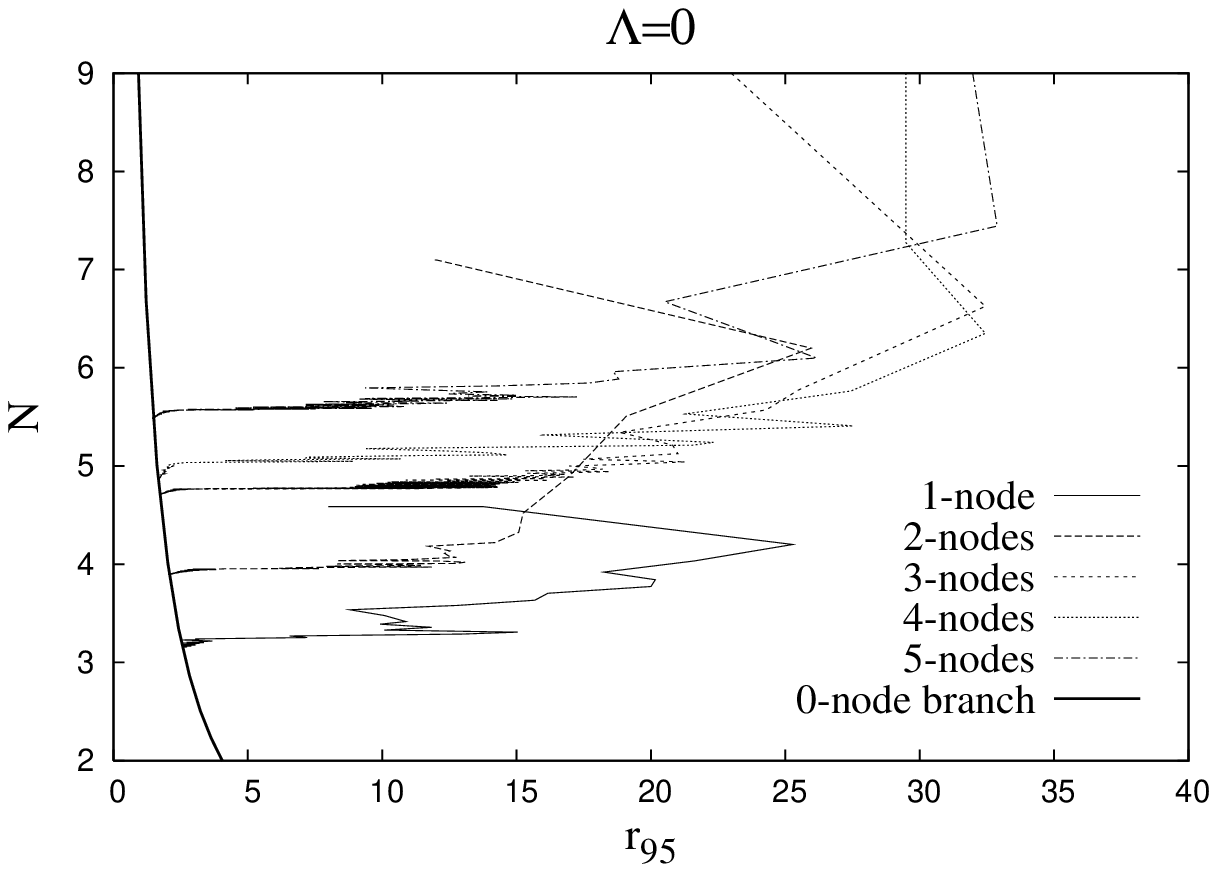}
\includegraphics[width=8cm]{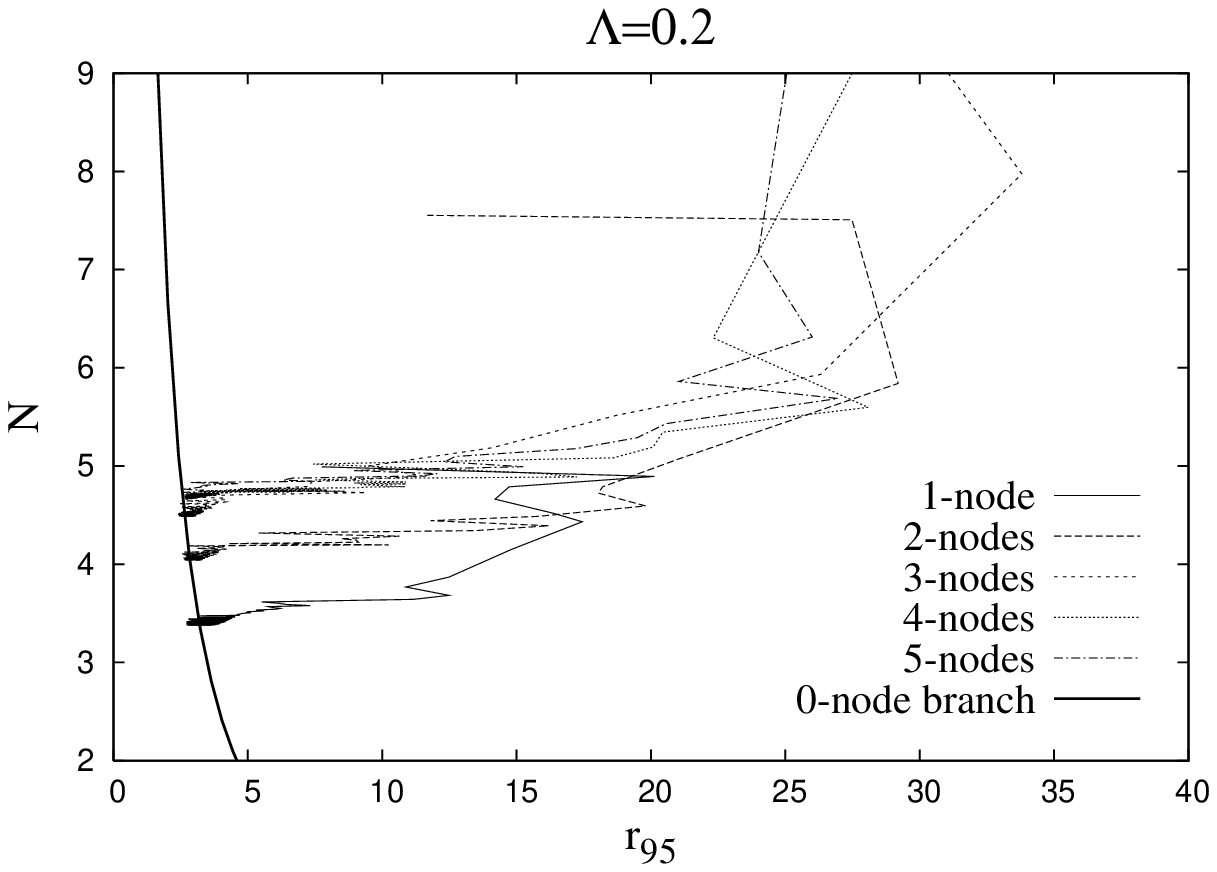}
\caption{\label{fig:excited} (Left) The evolution of initial equilibrium 
configurations in excited states with $\Lambda=0$. Such configurations 
start up with a given mass and radius (top-right in the figure) and after 
a while they approach one of the equilibrium solutions in the ground 
state through the emission of bursts of scalar field particles; the late 
time tendency is the approach to one of the points in the 
solid curve that contains all the possible ground state equilibrium 
configurations. This is a typical case of gravitational cooling. (Right). 
The same attractor behavior for $\Lambda=0.2$. In both plots the solid 
lines correspond to those shown in Figure \ref{fig:plotone}. For all 
these systems the quantity $2K+3I+W$ was oscillating around zero with an 
amplitude converging to zero.} 
\end{figure*}

To finish with, we show some numerical runs of excited states for 
$\Lambda=0$ and $\Lambda = 0.2$ in Fig.~\ref{fig:excited}. As expected, 
all excited configurations we tried are intrinsically unstable, and 
eventually settle down onto an equilibrium configuration. The initial 
excited equilibrium configurations start at the far top-right in 
the plots, and after a short time the systems approach the branches of 
ground state configurations shown in Fig.~\ref{fig:plotone}; the plot 
corresponding to the case $\Lambda=0$ is a reproduction of Fig. 16 in 
\cite{fsglau2004} with the appropriate redefined radius. The time 
of decay is of the order $10^2 m^{-1}$, which is a quite small time scale 
compared to the age of the universe, even for small values of the boson
mass\footnote{Even for an ultra-light boson mass of $10^{-23}$ eV,
the time decay is of order $\sim 100 m^{-1} \sim 10^3$ yr.}.

\section{Conclusions}
\label{sec:conclusions}

We have found stationary solutions to the spherically symmetric SP system 
of equations with a self-interaction term of the Gross-Pitaevskii type,
according to the mean field approximation of Bose condensates. We
found that equilibrium configurations exist that are virialized and stable
under the action of small perturbations. 

For initial profiles with a high kinetic energy, we have 
shown that there are bursts of scalar field involving a considerable 
amount of matter of the order of half the initial mass and for stationary 
excited initial configuration the emission of mass is even higher; after 
a while, the system relaxes and virializes around an equilibrium
configuration. This shows that equilibrium configurations are
late-time attractors. We have made numerical experiments with less
exotic initial profiles, like gaussians, and found the same tendency
to virialize and to accommodate around an equilibrium configuration.

What this shows is that a rather arbitrary initial fluctuation of 
self-gravitating scalar field, in the Newtonian regime, \emph{always}
evolves towards a virialized configuration, which should
    be compared to the relativistic case, in which some
    particular initial configurations lead to the formation of black
    holes, see\cite{seidel91,balakrishna98,alcubierre2003,guzman2004}. 
Hence, a
Bose condensate made of ultralight scalar field particles, as those
proposed to be the dark matter, {\it tolerates} the introduction of a
self-interaction term in the mean field approximation.

We expect the results presented all along this work could be
useful for models of scalar dark matter, as some papers have explored
the idea of including self-interaction terms 
\cite{matoslurena2004,arbeyetal,sahni2004}. We also hope this work
will help others to decide whether such objects can be found in the
cosmos.


\acknowledgments
This research is partly supported by the bilateral project DFG-CONACyT 
444-113/16/3-1; CONACyT grants 32138-E, 34407-E and 42748; PROMEP grants 
UGTO-CA-3 and UMICH-PTC-121; CIC-UMSNH-4.9 and Concyteg 05-16-K117-032. 
The runs were carried out in the Ek-bek cluster of the ``Laboratorio de 
Superc\'omputo Astrof\'{\i}sico (LASUMA)'' at CINVESTAV-IPN.


\end{document}